\documentclass[preprint,groupedaddress,nofootinbib,preprintnumbers,eqsecnum]{revtex4-1}

\usepackage{amsmath}
\usepackage{mathrsfs}
\usepackage[psamsfonts]{amssymb}
\usepackage{bm} 
\usepackage[dvipdfmx]{hyperref}
\newcommand{\g}{\gamma}  
\newcommand{\de}{\delta}  
\newcommand{\m}{\mu}  
\newcommand{\n}{\nu}  
 
\newcommand{\x}{\xi}

\newcommand{\p}{\psi}  
\newcommand{\pb}{\overline{\psi}}  
 
\newcommand{\vep}{\varepsilon}   
  
\newcommand{\ti}{\tilde}  

\newcommand{\Slash}[1]{{\ooalign{\hfil/\hfil\crcr$#1$}}}  
\begin{document}


\title{Heat kernel approach to the relations between covariant and consistent currents in chiral gauge theories}


\author{Masaharu Takeuchi}
\email{m.takeuchi.258@gmail.com}
\affiliation{Graduate  School  of  Science  and  Engineering, 
Yamagata  University,  Yamagata 990-8560, Japan}
\author{Ryusuke Endo}
\email{endo@sci.kj.yamagata-u.ac.jp}
\affiliation{Department of Physics, Yamagata University, Yamagata 990-8560, Japan}


\begin{abstract}
We apply the heat kernel method to 
relations between covariant and consistent currents in anomalous 
chiral gauge theories. 
Banerjee et al. have shown that the relation 
between these 
currents is expressed by a ``functional curl'' 
of the covariant current. 
Using the heat kernel method, we evaluate the functional curl explicitly in arbitrary even dimensions. 
We also apply the heat kernel method 
to evaluate 
Osabe and Suzuki's results of the difference between  covariant and consistent 
currents in two and four  dimensions. 
Applying the arguments of Banerjee et al. to 
gravitational anomalies, 
we investigate the relationship 
between the covariant and consistent energy--momentum tensors. 
The relation is found to be 
expressed by a functional curl of the covariant energy--momentum tensor. 
\end{abstract}

\preprint{YGHP-17-01}


\maketitle

%
%
%
%
\section{Introduction}
%
%
Chiral gauge anomalies can be viewed in one of two ways, 
namely, covariant and consistent. 
Covariant anomalies are defined as covariant divergences 
of the covariant current, i.e.,  a covariant divergence of the 
covariantly regularized  expectation value of the current.
Consistent anomalies can be considered as  gauge transformations of 
a regularized effective action. 
From this definition, consistent anomalies satisfy the the Wess--Zumino 
consistency condition \cite{Wess_Zumino_cons}.

The covariant and consistent anomalies are known to be equivalent in the sense that
they lead to the same anomaly-cancellation condition. 
Bardeen and Zumino \cite{Bardeen_Zumino_gauge_gravity} have given a general proof 
for this equivalence 
of the anomalies  using algebraic prescriptions. 
Their  approach does not need any explicit form for the Lagrangians, 
thus giving  model-independent results.
Lagrangian-based field-theoretical approaches to the equivalence of the gauge 
anomalies have been given by various authors 
\cite{Banerjee_Banerjee_Mitra, Banerjee_Banerjee, Banerjee_Banerjee_point_splitting,
      Fujikawa_Suzuki,Suzuki_SUSY,Osabe_Suzuki}. 
In particlular, 
 Banerjee et al.  \cite{Banerjee_Banerjee_Mitra} 
have shown  
equivalence by introducing  a regularized effective action 
defined through covariant current. 

To prove the equivalence of covariant and consistent gauge anomalies, 
Banerjee et al. \cite{Banerjee_Banerjee_Mitra} 
gave a relationship between the covariant and consistent currents. 
The consistent current was derived as a functional derivative of a regularized 
effective action, which was defined 
using the covariant current \cite{Banerjee_Banerjee_Mitra}. 
As a result, the relationship between the covariant and consistent currents is expressed by 
a ``functional curl'' of the 
covariant current\footnote{%
                  The functional curl of the covariant current  appears also
                  in the covariant commutator anomaly \cite{Dunne_Trugenberger,Kelnhofer}.
                         }. 
The authors of \cite{Banerjee_Banerjee_Mitra} argued that the functional curl of the 
covariant current is proportional to the delta function. 
With the help of the delta-function-type behavior of the functional curl, 
they have derived the relationship between the covariant and consistent gauge anomalies. 
Although their result agrees with Bardeen and Zumino \cite{Bardeen_Zumino_gauge_gravity}, 
the delta-function-type behavior of the functional curl is not clearly explained  in their arguments. 
Thus, it is desirable to prove the 
behavior 
of the functional curl more explicitly. 

The functional curl of the covariant current  has been discussed by various 
authors \cite{Grisaru_Penati, Dunne_Trugenberger, Kelnhofer, Fujikawa_Suzuki, 
              Suzuki_SUSY,  Z_Qiu_H_Ren}. 
Fujikawa and Suzuki \cite{Fujikawa_Suzuki} 
gave a formal proof of the relationship 
between the functional curl and the covariant anomaly; 
this relation was derived by Banerjee et al. \cite{Banerjee_Banerjee_Mitra} 
using 
the delta-function-type behavior of the functional curl. 
Ohshima et al. \cite{Suzuki_SUSY} 
evaluated the functional curl of the covariant current 
in supersymmetric chiral gauge theory. 
This curl was evaluated explicitly
by using the 
Fourier transformation in four dimensions. 
Based on their motivation which differed from that of Banerjee et al.,
Qiu and Ren \cite{Z_Qiu_H_Ren}  evaluated the functional curl 
explicitly by using the point-splitting method in two and four  dimensions. 
All of these results  are consistent with
the curl's delta-function-type behavior. 

Other studies concerning  the relationship  between the covariant and consistent currents 
have been reported 
in Refs. \cite{Banerjee_Banerjee, Banerjee_Banerjee_point_splitting, Osabe_Suzuki}, where 
the functional curl does not appear 
in the arguments.  
The difference between the covariant and consistent currents has been directly 
calculated using 
Pauli--Villars regularization \cite{Banerjee_Banerjee}
and 
the point-splitting method \cite{Banerjee_Banerjee_point_splitting}. 
Osabe and Suzuki \cite{Osabe_Suzuki}  
also discussed the difference between  covariant and consistent currents, 
which they defined 
by invoking  
different types of exponential regulators. 
These regulators were then used to obtain a formal expression of the difference 
between the covariant and consistent currents. 

In this paper, 
by using the heat kernel method \cite{DeWitt_HK}, we evaluate the functional curl of the covariant current explicitly. 
The  curl that we derive agrees with that of 
Refs. \cite{Banerjee_Banerjee_Mitra, Fujikawa_Suzuki}.
Our result presents another direct proof of the delta-function-type 
behavior of the functional curl 
in arbitrary even dimensions. 
We also  apply the heat kernel method  to evaluate  
Osabe and Suzuki's formal expression of 
the difference between  the covariant and consistent currents \cite{Osabe_Suzuki}.   
This difference, which we calculate in two and four  dimensions, agrees with 
the previous results 
 \cite{Bardeen_Zumino_gauge_gravity,Banerjee_Banerjee_Mitra}. 
The arguments of Banerjee et al. \cite{Banerjee_Banerjee_Mitra} are also 
applied to  gravitational anomalies \cite{Alvarez-Gaume_Witten}\footnote{%
              The equivalence of the covariant and consistent gravitational 
              anomalies is also shown in Ref.\cite{Bardeen_Zumino_gauge_gravity} 
              by the algebraic approach. We are interested here in the field theoretical 
              approach to the equivalence. 
                   }. 
We investigate the relationship 
 between the covariant and consistent energy--momentum tensors, 
which is found to be expressed by a functional curl of the covariant 
energy--momentum tensor.

The the rest of this paper is outlined as follows.
In Sect. \ref{2}, 
we review the arguments of Banerjee et al. \cite{Banerjee_Banerjee_Mitra} 
concerning  covariant and consistent gauge anomalies.  
In Sect. \ref{3}, 
we evaluate the functional curl of the covariant current explicitly by using the heat kernel method in 
arbitrary even dimensions. 
In Sect. \ref{4}, 
we apply the heat kernel method  to Osabe and Suzuki's difference of the 
covariant and consistent currents \cite{Osabe_Suzuki}  in two and four dimensions.
In Sect. \ref{5}, 
by applying the arguments of Banerjee et al. \cite{Banerjee_Banerjee_Mitra} 
to the gravitational anomalies, 
we investigate the relationship between the covariant and consistent 
energy--momentum tensors.
Section \ref{6} 
is devoted to a summary and discussion.
%
%
%
%
%
%
\section{Functional curl of the covariant current}
\label{2}
We consider a chiral gauge theory 
given by the following $2n$-dimensional Euclidean Lagrangian 
\begin{equation} 
   \mathscr{L} 
   = \overline{\psi} \gamma^\mu (\partial_\mu + iA_\mu^a T^a \frac{1 - \gamma_5}{2}) \psi , 
\label{eq:lagrangian}
\end{equation}
where $\psi$ and $\overline{\psi}$ are the Dirac spinors, 
and $A_\mu^a$ are the gauge fields. 
The metric we use is $\eta^{\mu \nu} = - \delta^{\mu \nu}$. 
The Dirac gamma matrices $\gamma^\mu$ are anti-hermitian, 
and  $\gamma_5 = i^n \g^1 \g^2 \cdots \g^{2n}$ is hermitian. 
The matrices $\gamma^\mu$ and the hermitian generators $T^a$ satisfy
\begin{align}
   &\{ \g^{\mu},\g^{\nu}\} = - 2 \de^{\m \nu} , 
   \label{eq:uc} 
   \\
   &[T^a,T^b] = if^{abc}T^c , 
   \\
   &\text{tr} T^a T^b  = \frac{1}{2}\delta^{ab} , 
   \label{eq:normalization}
\end{align}
where $f^{abc}$ are the structure constants of the gauge group. 
The Lagrangian $\mathscr{L}$ is invariant under these gauge transformations: 
\begin{gather}
    \delta_\alpha \p (x) = i \alpha^a (x) T^a \frac{1 - \g_5}{2} \p (x) , 
    \\        
    \delta_\alpha \pb (x) = - i \pb (x) \alpha^a(x) T^a \frac{1 + \g_5}{2} , 
    \\ 
    \delta_\alpha A_\mu^a (x) = - D_\mu \alpha^a (x)  
                         = - \partial_\mu \alpha^a (x) + f^{abc}A_\mu^b(x) \alpha^c (x) .
\label{eq:gaugetr3}
\end{gather}
%
%
%
%
\subsection{Covariant and consistent currents}
Although Lagrangian \eqref{eq:lagrangian} is invariant under gauge transformations, 
the effective action is not. The effective action 
$W[A_\mu^a]$ transforms as 
\begin{align}
     \delta_\alpha W[A_\mu^a] 
     &= \int dx \frac{\delta W}{\delta A_\mu^a(x)} \delta_\alpha A_\mu^a(x) 
     \notag 
     \\
     &= i\int dx \alpha^a(x)G^a(x) ,
\end{align}
where the gauge anomaly $G^a(x)$ is defined by
\begin{equation}
     G^a(x) = D_\mu \langle J^{\mu a}(x) \rangle 
   \label{eq:Wtr}
\end{equation}
with the vacuum expectation value of the current $\langle J^{\mu a}(x) \rangle$ 
given by 
\begin{equation}
  \langle J^{\mu a}(x) \rangle 
  = \frac{1}{i} \frac{\delta}{\delta A_\mu^a (x)} W[A_\nu^b] . 
\label{eq:vcurrent}
\end{equation}
These expressions have only formal meanings, i.e., 
$\langle J^{\mu a}(x) \rangle$ is divergent since $W[A_\nu^b]$ 
is divergent. 
To treat  current \eqref{eq:vcurrent} meaningfully, 
we should adopt an appropriate regularization.

We usually adopt either 
the consistent or  covariant regularization.
The consistently regularized current 
$\langle J^{\mu a}(x) \rangle_{\text{cons}}$ is defined through the regularization of 
the effective action, $W[A_\nu^b]$. 
Using the regularized 
effective action $W_{\text{reg}}[A_\nu^b]$, we define 
a regularized current 
\begin{equation}
    \langle J^{\mu a}(x) \rangle_{\text{cons}} 
    = \frac{1}{i} \frac{\delta}{\delta A_\mu^a (x)} W_{\text{reg}}[A_\nu^b] .
\label{eq:conscurrent}
\end{equation}
We note that the consistent current 
$\langle J^{\mu a}(x) \rangle_{\text{cons}}$ 
given by \eqref{eq:conscurrent} satisfies 
the integrability condition:
\begin{equation} 
    \frac{\delta}{\delta A_\mu^a (x)} \langle J^{\nu b}(x') \rangle_{\textrm{cons}} 
  - \frac{\delta}{\delta A_\nu^b (x')} \langle J^{\mu a}(x) \rangle_{\textrm{cons}} 
  = 0 .  
\label{eq:integrability}
\end{equation}
If 
$W_{\text{reg}}[A_\nu^b]$ is gauge 
invariant, the current 
$\langle J^{\mu a}(x) \rangle_{\text{cons}}$ 
transforms covariantly under gauge transformation.
In the anomalous gauge theory, however,  $W_{\text{reg}}[A_{\nu}^{b}]$ is 
not gauge invariant and thus the current 
$\langle J^{\mu a}(x) \rangle_{\text{cons}}$ does not transform covariantly.

The covariant current, $\langle J^{\mu a}(x) \rangle_{\text{cov}}$, 
is the expectation 
value of current regularized 
covariantly with respect to gauge transformation. 
In contrast with the current $\langle J^{\mu a}(x) \rangle_{\text{cons}}$, 
$\langle J^{\mu a}(x) \rangle_{\text{cov}}$
transforms covariantly under the gauge transformation \eqref{eq:gaugetr3}. 
Consequently, 
$\langle J^{\mu a}(x) \rangle_{\text{cov}}$ cannot be expressed 
in the form 
of \eqref{eq:conscurrent} in the anomalous theory. 
In particular, 
the covariant current does not satisfy the integrability 
condition \eqref{eq:integrability}. 
These expectation values are functionals of $A_\nu^b$.
When we need to pay attention to the functional property, we use a symbol such as 
$\langle J^{\mu a}(x) \rangle_{\text{cov}}[A_\nu^b]$.

Substituting these regularized currents into equation \eqref{eq:Wtr}, we obtain 
the following gauge anomalies: 
\begin{equation}
  G^a_{\textrm{cov}}(x) = D_\mu \langle J^{\mu a}(x) \rangle_{\textrm{cov}}
\end{equation}
and 
\begin{equation}
  G^a_{\textrm{cons}}(x) = D_\mu \langle J^{\mu a}(x) \rangle_{\textrm{cons}} ,
\end{equation}
where $G^a_{\textrm{cov}}(x)$ and $G^a_{\textrm{cons}}(x)$ are called 
covariant and consistent, 
respectively. 
The consistent anomaly, $G^a_{\textrm{cons}}(x)$, satisfies the Wess--Zumino 
consistency condition
 \cite{Wess_Zumino_cons}, which is ascribed to the integrability 
condition \eqref{eq:integrability}.

The covariant anomaly $G^{a}_{\text{cov}}(x)$ 
can be expressed as (see, for example, \cite{Fujikawa_Suzuki})
\begin{equation}
  G^a_{\textrm{cov}}(x) 
    = \lim_{s \to 0} \lim_{x^{\prime} \to x} \text{tr} 
         T^a \gamma_5 e^{ - s\not D^2} \delta (x - x') , 
\label{eq:covanomaly}
\end{equation}
where $s$ is the cut-off parameter and 
$\Slash{D} = \g^{\mu}(\partial_{\mu} + iA_{\mu}^{a}T^{a})$. 
This quantity can be calculated \cite{Fujikawa_Suzuki} 
as 
\begin{equation}
  G^a_{\textrm{cov}}(x) 
     = \frac{ (-1)^n}{(4\pi )^{n} n!} \varepsilon^{\mu_1 \nu_1 \dotsm \mu_n \nu_n}
                        \text{tr} T^a F_{\mu_1 \nu_1}(x) \dotsm F_{\mu_n \nu_n}(x) , 
\label{eq:covanomalyfinal}
\end{equation}
where $\varepsilon^{\mu_{1}\nu_{1}\dotsm \mu_{n}\nu_{n}}$ 
is the totally antisymmetric tensor with 
$\varepsilon^{12 \dotsm 2n} = 1$ 
and $F_{\m \nu} = \partial_{\mu}A_{\nu} - \partial_{\nu}A_{\mu} + i[A_{\mu},A_{\nu}]$ is 
the field strength of the gauge fields $A_{\mu} = A_{\mu}^{a}T^{a}$. 
The covariant anomaly $G^{a}_{\textrm{cov}}(x)$ 
is a finite local polynomial of field strength $F_{\m \nu}(x)$.
%
%
%
\subsection{Relationship between the covariant and consistent currents}
We follow Banerjee et al. \cite{Banerjee_Banerjee_Mitra} in  deriving 
the relationship between the covariant and consistent currents. 
We introduce a parameter $g$ and define
\begin{equation}
 W_g  = W[gA_\mu^a] . 
\end{equation}
If we put $g = 1$, $W_g$ reduces to the original effective action $W[A_\mu^a]$.
We can express  $W[A_\mu^a ]$  using $W_g $ as
\begin{equation}
   W[A_\mu^a] = \int_0^1 dg \frac{\partial W_g}{\partial g} + W_{g = 0} .
\end{equation}
Note that the $g$-dependence arises only through the combination $gA_\mu^a$, we obtain
\begin{equation}
   W[A_\mu^a] 
      = \int_0^1 dg \int  dx\, 
         \frac{\delta W_g }{\delta \bigl( gA_\mu^a (x) \bigr)}  A_\mu^a (x) , 
\label{eq:W}
\end{equation}
where we have dropped the $W_{g = 0}$ term since it is an $A_\mu^a$-independent constant.
From definition \eqref{eq:vcurrent}, we rewrite \eqref{eq:W} as 
\begin{equation}
  W[A_\mu^a] = i \int_0^1 dg \int  dx\, A_\mu^a (x)\langle J^{\mu a}(x) \rangle^g ,  
\label{eq:beforeregW}
\end{equation}
where we have used the notation
\begin{equation}
\langle J^{\mu a}(x) \rangle^g = \langle J^{\mu a}(x) \rangle [gA_{\nu}^{b}] .
\end{equation}
Expression \eqref{eq:beforeregW} has only a formal meaning 
because the current $\langle J^{\mu a}(x) \rangle^g$ 
is not yet regularized.  
The crucial point of the prescription of 
Ref. \cite{Banerjee_Banerjee_Mitra} is to substitute 
covariant current 
$
  \langle J^{\mu a}(x) \rangle^g_{\text{cov}} 
    = \langle J^{\mu a}(x) \rangle_{\text{cov}} [gA_\nu^b]
$ 
for 
$\langle J^{\mu a}(x) \rangle^{g}$ 
in \eqref{eq:beforeregW} to construct a regularized effective action,  
$W[A_\mu^a]_{\text{reg}}$: 
\begin{equation}
  W[A_\mu^a]_{\text{reg}} 
        = i \int_0^1 dg \int  dx\, A_\mu^a (x)
                             \langle J^{\mu a}(x) \rangle^g_{\text{cov}} .  
\label{eq:W_reg_cov_current}
\end{equation}
We can obtain a consistent current from the regularized effective 
action \eqref{eq:W_reg_cov_current}.
Taking the functional derivative of \eqref{eq:W_reg_cov_current} 
with respect to $A_\mu^a (x)$, we obtain the relationship 
between the covariant and consistent currents \cite{Banerjee_Banerjee_Mitra}:
\begin{equation}
  \langle J^{\mu a}(x) \rangle_{\text{cons}} 
       = \langle J^{\mu a}(x) \rangle_{\text{cov}}
        + \int_0^1 dg \int dx'  g A_\nu^b (x')
              \left\{ 
                     \frac{\delta \langle J^{\nu b}(x') \rangle^{g}_{\text{cov}}}
                          {\delta \left( gA_\mu^a (x) \right) }
                   - \frac{\delta \langle J^{\mu a}(x) \rangle^g_{\text{cov}} }
                          {\delta \left( gA_\nu^b (x') \right) } 
              \right\} . 
\label{eq:rerationaleq}
\end{equation} 
Note  that the ``functional curl'' of the covariant current appears 
in the integrand of the second term on the 
right-hand side. 
The functional curl in \eqref{eq:rerationaleq} is obtained 
by substituting $gA_\mu^a$ into $A_\mu^a$ in 
the functional curl
\begin{equation}
       \frac{\delta \langle J^{\nu b}(x') \rangle_{\text{cov}}} 
            {\delta A_\mu^a (x)}
     - \frac{\delta \langle J^{\mu a}(x) \rangle_{\text{cov}} } 
            {\delta A_\nu^b (x')} ,
\end{equation}  
which does not vanish since the covariant current does not satisfy 
the integrability condition \eqref{eq:integrability} in 
the anomalous theory.\footnote{%
            It can be seen that the parity-conserving part  of the 
            functional curl vanishes. 
                              }  
Taking the covariant divergence of \eqref{eq:rerationaleq}, we obtain the relationship 
between the covariant and consistent 
gauge anomalies:
\begin{equation}
  G^a _{\text{cons}}(x) 
      = G^a _{\text{cov}}(x) 
      + D_{\mu}^{ac}\! \int_0^1 dg\int dx'  gA_{\nu}^b(x' )
             \left\{
                 \frac{\delta \langle J^{\nu b}(x') \rangle^g_{\text{cov}}}
                      {\delta \left( gA_\mu^c (x) \right) }
               - \frac{\delta \langle J^{\mu c}(x) \rangle^g_{\text{cov}} }
                      {\delta \left( gA_\nu^b (x') \right) }  
             \right\} , 
\label{eq:currentfc}
\end{equation} 
where $D_{\mu}^{ac} = \de^{ac}\partial_{\mu} - f^{adc}A_{\nu}^{d}$. 

Banerjee et al. \cite{Banerjee_Banerjee_Mitra} have evaluated the functional curl 
of the covariant current by using the fact 
that this curl has delta-function-type behavior at $x = x' $: 
\begin{equation}
         \frac{\delta \langle J^{\nu b}(x') \rangle_{\text{cov}}}
              {\delta A_\mu^a (x)}
       - \frac{\delta \langle J^{\mu a}(x) \rangle_{\text{cov}} }
              {\delta A_\nu^b (x')}  
        \propto 
                \delta (x - x' ) . 
\label{eq:assumption}
\end{equation}
Using \eqref{eq:assumption}, they showed that the functional curl can be 
expressed by the covariant gauge anomaly, 
\begin{equation}
         \frac{\delta \langle J^{\nu b}(x' ) \rangle_{\text{cov}}}
              {\delta A_\mu^a (x)}
       - \frac{\delta \langle J^{\mu a}(x) \rangle_{\text{cov}}}
              {\delta A_{\nu}^b(x' )} 
       = 
         - 2\frac{\delta G^b_{\text{cov}}(x' )}
                 {\delta F_{\mu \nu }^a(x)} . 
\label{eq:fc-covanomaly}
\end{equation} 
Substituting this equation into \eqref{eq:currentfc}, 
they  derived an expression for the consistent gauge anomaly 
that agrees with the result of Ref. \cite{Bardeen_Zumino_gauge_gravity}. 
In the arguments of Ref. \cite{Banerjee_Banerjee_Mitra} given above, 
it is crucial for equation \eqref{eq:assumption} to actually hold. 
In Ref. \cite{Banerjee_Banerjee_Mitra}, however, 
a detailed proof of \eqref{eq:assumption} is not shown. 
Considering this point, we evaluate the functional curl 
explicitly in the next section.
%
%
%
%
%
%
%
%
%
\section{Explicit evaluation of the functional curl of the covariant current}
\label{3}
The expectation value of the current can be expressed by
\begin{align}
   \langle J^{\mu a}(x) \rangle 
     &= \Big{\langle} 
           \overline{\psi}(x)\gamma^\mu T^a \frac{1 - \gamma_5}{2} \psi(x)
        \Big{\rangle} 
     \notag 
     \\
     &= \lim_{x'  \to x} \text{tr} 
              \frac{1 + \gamma_5}{2} \gamma^\mu T^a 
              \frac{1}{{\Slash{D}}} \delta (x - x') , 
\label{eq:original_current}
\end{align}
where $\Slash{D} = \gamma^\mu (\partial_\mu + iA_\mu^a T^a)$.
To regularize \eqref{eq:original_current}, 
we employ the Gaussian regulator to define a covariant current \cite{Fujikawa_Suzuki},
\begin{equation}                                    
       \langle J^{\mu a}(x) \rangle_{\textrm{cov}} 
        = \lim_{s \to 0} \lim_{x'  \to x} \text{tr}
             \frac{1 + \gamma_5}{2} \gamma^{\mu} T^a \frac{1}{{\Slash{D}}}
              e^{ - s \not D^2 }  \delta (x - x') , 
\label{eq:covariant-current-reg}
\end{equation}
where $s$ is the cut-off parameter. 
Because the regulator $e^{ - s\not D^2 }$ is covariant, 
the current $ \langle J^{\mu a}(x) \rangle_{\textrm{cov}}$ 
transforms covariantly. 
Taking the functional curl of \eqref{eq:covariant-current-reg} 
and using trace properties, 
we have \cite{Fujikawa_Suzuki}
\begin{align}
  & \frac{\delta \langle J^{\nu b}(x') \rangle_{\text{cov}}}
         {\delta A_\mu^a (x)} 
  - \frac{\delta \langle J^{\mu a}(x) \rangle_{\text{cov}}}
         {\delta A_\nu^b(x' )} 
  \notag 
  \\
  &= 
   - i \lim_{s \to 0} s\, \text{tr} \gamma_5 \gamma^\nu T^b \int_0^1 d\alpha 
            \left( 
                e^{ - (1 - \alpha)s \not D^{\prime 2}} \delta (x' - x )
            \right) 
            \gamma^\mu T^a e^{- \alpha s \not D^2} \delta (x - x') , 
\label{eq:covfc} 
\end{align}
where 
$
\Slash{D}^{\prime} = \gamma^\mu (\partial_\mu^\prime + iA_\mu^a(x') T^a)
$.
Here Fujikawa and Suzuki have shown that the right-hand side of \eqref{eq:covfc} 
is equal to the functional derivative 
of the expression for the covariant anomaly \eqref{eq:covanomaly} with respect 
to the field strength \cite{Fujikawa_Suzuki}, 
which gives a formal proof of \eqref{eq:fc-covanomaly} and thus gives the 
proof of \eqref{eq:assumption}. 
%

In the following, we evaluate the functional curl \eqref{eq:covfc} 
explicitly by using the heat kernel method \cite{DeWitt_HK}.
The functional curl \eqref{eq:covfc} can be expressed by
\begin{align}
   &\frac{\delta \langle J^{\nu b}(x') \rangle_{\text{cov}}}
         {\delta A_\mu^a (x)} 
  - \frac{\delta \langle J^{\mu a}(x) \rangle_{\text{cov}}}
         {\delta A_{\nu}^b(x' )} 
   \notag 
   \\
   &= - i \lim_{s \to 0} s\, \text{tr} \gamma_5 \gamma^\nu T^b
                 \int_0^1  d\alpha K(x',x;(1 - \alpha )s) 
                 \gamma^\mu T^a K(x,x' ;\alpha s) , 
\label{eq:fc1}
\end{align}
where $K(x,x' ;s)$ is the heat kernel defined by 
\begin{equation}
      K(x,x';s) = e^{ - s\not D^2 }\delta (x - x' ) .
\end{equation}
Substituting the heat-kernel expansion
\begin{equation}
    K(x,x';s) 
    = \frac{1}{(4\pi s)^n }  e^{(x - x' )^2 /4s} \sum_{k = 0}^\infty a_k (x,x' ) s^k 
\label{eq:K-expansion-0}
\end{equation}
into equation \eqref{eq:fc1}, we have
\begin{align}
     \eqref{eq:fc1} 
     = - i\frac{1}{(4\pi )^{2n}}  \int_0^1 d\alpha  \sum_{k,l}
        &(1 - \alpha )^{k - n}\alpha^{l - n} s^{1 + k + l - 2n} 
                                     e^{(x - x' )^2 /4\alpha(1 - \alpha)s} 
     \notag \\
        &\times \text{tr} \gamma_5 \gamma^\nu T^b a_{k}(x' ,x) \gamma^\mu T^a a_{l}(x,x' ) ,  
\label{eq:fc2}
\end{align}
where we have suppressed the symbol $\displaystyle \lim_{s \to 0}$.
The exponential function appearing on the right-hand side can be understood 
as the heat kernel of the free theory. 
That is, if we define
\begin{equation}
           K_0 (x,x';s) =  \frac{1}{(4\pi s)^n} e^{(x - x' )^2 /4s} , 
\label{eq:solution1}
\end{equation}
then $K_0 (x,x';s)$ satisfies
\begin{equation}
      \frac{\partial}{\partial s} K_0(x,x';s) 
      = - \square K_0(x,x';s) ,         
      \quad   K_0(x,x';s = 0) = \delta (x - x') ,
\label{eq:solution}
\end{equation}
where $\square = \partial_\mu \partial^\mu$.
A formal solution to \eqref{eq:solution} can be written as
\begin{equation}
   K_0(x,x';s) =  e^{ - s\square} \delta (x - x') .
\label{eq:solution2}
\end{equation}
Taking the Taylor expansion of $e^{ - s\square}$ with respect to $s$, 
we have \footnote{A proof of \eqref{eq:freehk} 
                  using test functions is given in Appendix \ref{C}.
                  }
\begin{equation}
   e^{(x - x' )^2 /4s} 
   = (4\pi s)^n \sum_{k = 0}^\infty \frac{( - s\square )^{k}}{k!} \delta (x - x' ) .
\label{eq:freehk}
\end{equation}
With this formula and the integration formula 
\begin{equation}
    \int_0^1 d\alpha (1 - \alpha )^{k + m} \alpha^{l + m} 
                           = \frac{(k + m)!(l + m)!}{(k + l + 2m +1)!} ,
\label{eq:beta}
\end{equation}
equation \eqref{eq:fc2} can be written as
\begin{align}
   \eqref{eq:fc2} 
   = - i\frac{1}{(4\pi )^n } \sum_{k,l,m} 
           & \frac{1}{m!} \frac{(k + m)!(l + m)!}{(k + l + 2m +1)!} s^{1 + k + l + m - n} 
           \notag \\
           &\times \text{tr} \gamma_5 \gamma^\nu T^b a_{k}(x' ,x)
                \gamma^\mu T^a a_{l}(x,x' )(- \square)^m \delta (x - x' ) , 
\label{eq:fc4}
\end{align}
Considering that the terms higher than $0$-th order in $s$ vanish in the limit $s \to 0$,
we find that the indices $k$, $l$, and $m$ of the surviving terms \eqref{eq:fc4} 
satisfy the condition 
\begin{equation}
       1 + k + l + m - n \le 0 .          \label{eq:cond1}
\end{equation}
In addition, the surviving terms must contain at least $2n$ factors of 
gamma matrices $\gamma^\mu$, because of the existence 
of $\gamma_5$ in the trace over spinor indices. 
As shown in Appendix \ref{A}, $a_{k}(x,x' )$ contains at most $2k$ factors 
of $\gamma^\mu$. 
Consequently,  indices $k$ and $l$ of the surviving terms satisfy the condition
\begin{equation}
             2 + 2k +2l \ge 2n .      \label{eq:cond2}
\end{equation}
Conditions \eqref{eq:cond1} and  \eqref{eq:cond2} lead to
\begin{align}
    &m = 0 , 
    \\
    &l = n - k -1 .
\end{align}
Then, \eqref{eq:fc4} becomes
\begin{align}
& \eqref{eq:fc4} 
   \notag 
   \\
   &= - i\frac{1}{(4\pi )^n } \sum_{k = 0}^{n - 1}
         \frac{k!(n - k - 1)!}{n!}
         \text{tr} \gamma_5 \gamma^\nu T^b a_{k}(x,x)\gamma^\mu T^a 
a_{n - k - 1}(x,x)\de (x - x' )  .
    \label{eq:fc5_pr}
\end{align}
$a_{k}(x,x)$ is given by \eqref{eq:gammamax1}, 
starting with the term containing  $2k$ factors of $\gamma^\mu$:
\begin{equation}
  a_{k}(x,x) 
   = \frac{(-1)^{k}}{k!} 
          \left( 
                 \frac{i}{2}\gamma^\mu \gamma^\nu F_{\m \nu} 
          \right)^{k} 
          + \cdots , 
    \tag{\ref{eq:gammamax1}}
\end{equation}
where the dots on
the right-hand side express terms with lower power of $\gamma^\mu$. 
Substituting  \eqref{eq:gammamax1} into \eqref{eq:fc5_pr}, 
we obtain the final expression for the functional curl, 
\begin{align}
   &\frac{\delta \langle J^{\nu b}(x' ) \rangle_{\text{cov}}}
         {\delta A_\mu^a (x)} 
  - \frac{\delta \langle J^{\mu a}(x) \rangle_{\text{cov}}}
         {\delta A_{\nu}^b(x' )} 
   \notag 
   \\
   &= - 2\frac{(-1)^n }{(4\pi )^n (n - 1)!} 
         \vep^{\m \n \m_1  \nu_1 \cdots \mu_{n - 1}\n_{n - 1}}
         \text{Str} T^a T^b F_{\m_1 \n_1 }\cdots F_{\m_{n - 1}\n_{n - 1}}
         \delta (x - x' ) , 
\label{eq:finalexfc}
\end{align}
where the symbol ``Str'' denotes the  symmetrized trace \cite{Zumino_Y_S} 
indicating  that the 
factors in the trace are to be totally symmetrized. We notice here that our 
evaluation gives a direct proof 
of \eqref{eq:assumption}.
Comparing this expression with the final expression of the covariant 
anomaly \eqref{eq:covanomalyfinal}, 
we again obtain  \eqref{eq:fc-covanomaly}. 
%
%
%
%
%
%
%
\section{Explicit evaluation of Osabe and Suzuki's expression for the current difference}
\label{4}
Osabe and Suzuki \cite{Osabe_Suzuki} have also discussed the difference between
 the consistent and covariant currents. 
Their consistent current, $\langle J^{\mu a}(x) \rangle_{\textrm{cons}}$, 
 can be written as 
\begin{equation}
  \langle J^{\mu a}(x) \rangle_{\textrm{cons}} 
    = \lim_{s \to 0} \text{tr} 
             \Big{\langle}x\Big{|} 
                      \frac{1 + \gamma_5}{2} \gamma^\mu T^a 
                            \frac{1}{{\Slash{D}}}
                            e^{ - s\not D \not{\,\partial}}
             \Big{|}x\Big{\rangle}
\label{eq:consistent-current-reg}
\end{equation}
in our notation, 
while the covariant current, $\langle J^{\mu a}(x) \rangle_{\textrm{cov}}$, is given 
by \eqref{eq:covariant-current-reg}, i.e., 
\begin{equation}                                   
  \langle J^{\mu a}(x) \rangle_{\textrm{cov}} 
    = \lim_{s \to 0} \text{tr}
             \Big{\langle}x\Big{|}
                      \frac{1 + \gamma_5}{2} \gamma^\mu T^a 
                      \frac{1}{\Slash{D}} e^{ - s \not D^2 }
             \Big{|}x\Big{\rangle} .
\end{equation}
From these definitions, they  derived an expression for the difference between currents. 
Their derivation can be 
explained essentially as follows: 
Introducing 
$\Slash{D}_{g} = \gamma^\mu D_{\mu}^{g} = \gamma^\mu (\partial_{\mu} + igA_{\mu})$ 
and noticing the equality 
\begin{equation}
     e^{ - s \not D^2} 
     - e^{ - s\not D \not{\,\partial}}
    = \int_0^1 dg 
        \frac{d}{dg} e^{ - s\not D\not D_{g}} ,
\end{equation}
we obtain
\begin{align}
 & \langle J^{\mu a}(x) \rangle_{\textrm{cons}} 
 - \langle J^{\mu a}(x) \rangle_{\textrm{cov}} 
 \notag 
 \\
 &= - \lim_{s \to 0} \int_0^1 dg \frac{d}{dg} \text{tr}
           \Big{\langle}x\Big{|} 
                   \frac{1+\gamma_5}{2} \gamma^\mu T^a 
                   \frac{1}{{\Slash{D}}} 
                   e^{ - s\not D\not D_{g}}
           \Big{|}x\Big{\rangle} 
 \notag 
 \\
 &= - \lim_{s \to 0} \int_0^1 dg \int_0^1 d\alpha \, \text{tr}
        \Big{\langle}x\Big{|} 
            \frac{1+\gamma_5}{2} \gamma^\mu T^a 
            \frac{1}{{\Slash{D}}}
            e^{ - (1 - \alpha )s\not D\not D_{g}}
            \left( 
                  - s\Slash{D}\frac{d\Slash{D}_{g}}{dg} 
            \right) 
            e^{ - \alpha s\not D\not D_{g}}
         \Big{|}x\Big{\rangle} 
  \notag 
  \\
  &= \lim_{s \to 0} s \int_0^1 dg \int_0^1 d \alpha \, \text{tr}
            \frac{1+\gamma_5}{2} \gamma^\mu T^a 
            \Big{\langle} x\Big{|}
                e^{ - (1-\alpha)s\not D_{g}\not D}
                \frac{d\Slash{D}_{g}}{dg}
                e^{-\alpha s \not D\not D_{g}}
            \Big{|}x\Big{\rangle} .
\label{eq:ourderivation}
\end{align}
In the third line, we have used the identity: 
\begin{equation}
     e^{ - s\not D \not D_g} {\Slash{D}} 
        = {\Slash{D}} e^{ - s\not D_g \not D} .
\end{equation}
Equation \eqref{eq:ourderivation} is equivalent to Osabe and Suzuki's 
expression for the current difference
 \cite{Osabe_Suzuki}.
%

Now, we calculate  current difference \eqref{eq:ourderivation} 
by applying the heat kernel method. 
Introducing heat kernels
\begin{align}
  & K_g(x,x' ;s) 
      = \langle x|e^{ - s \not D \not D_g}|x' \rangle , 
    \\
  &\tilde{K}_g(x,x' ;s) 
      = \langle x|e^{ - s \not D_g \not D}|x' \rangle ,
\end{align}
we express \eqref{eq:ourderivation} as 
\begin{align}
  &\langle J^{\mu a}(x) \rangle_{\textrm{cons}} 
  - \langle J^{\mu a}(x) \rangle_{\textrm{cov}} 
    \notag 
    \\
    &= \frac{i}{2} \lim_{s \to 0} s \int_0^1 dg \int_0^1 d\alpha \int dx' 
       \text{tr} \gamma_5 \gamma^\mu T^a \ti{K}_g(x,x';(1-\alpha)s) 
                                 \Slash{A}^\prime K_g (x',x;\alpha s) ,
\label{eq:diffK}
\end{align}
where $\Slash{A}^\prime = \gamma^\nu A_\nu (x')$ 
and we have omitted the parity-conserving terms since only 
parity-violating terms contribute to the anomalies.
These kernels $K_g$ and $\tilde{K}_g$ are not independent of  each other. 
In fact,  owing to the relation
\begin{equation}
     \Slash{D}_{g}\Slash{D} 
         = \left( 
                  \Slash{D}\Slash{D}_{g} 
           \right)^\dagger ,
\end{equation}
they satisfy
\begin{equation}
        \tilde{K}_g(x,x';s) = K_g(x',x;s)^\dagger .
            \label{eq:Kdagger}
\end{equation}
We expand $K_g(x,x';s)$ and $\tilde{K}_g(x,x';s)$ in $2n$ dimensions as follows:
\begin{align}
      &K_g(x,x' ;s) 
         = \frac{1}{(4\pi s)^n} e^{(x - x')^2/4s}
               \sum_{k = 0}^\infty b_k(x,x')s^k
      \label{eq:Ktildeexpand} , 
      \\
      &\tilde{K}_g(x,x';s) 
         = \frac{1}{(4\pi s)^n } e^{(x - x')^2/4s}
               \sum_{k = 0}^\infty \tilde {b}_k(x,x')s^k . 
      \label{eq:Kexpand}                    
\end{align}
Note here that \eqref{eq:Kdagger} indicates 
\begin{gather}
          \ti{b}_k(x,x' ) = b_k(x' ,x)^{\dagger} . 
\label{eq:tilde-dagger}
\end{gather}
Substituting expansions \eqref{eq:Ktildeexpand} and \eqref{eq:Kexpand} 
into \eqref{eq:diffK}, we have
\begin{align}
  \eqref{eq:diffK} 
    = \frac{i}{2} \frac{1}{(4\pi )^{2n}} \int_0^1 & dg\int_0^1 d\alpha \int dx' 
         \sum_{k,l} (1- \alpha)^{k - n} \alpha^{l- n} s^{1 + k + l - 2n} 
    \notag 
    \\
         &\times 
         e^{(x - x')^2/4\alpha (1-\alpha)s} \text{tr} \gamma_5 \gamma^\mu T^a 
                    \ti{b}_k(x,x' )\Slash{A}^\prime b_l(x',x) , 
\label{eq:diffexpand}
\end{align}
where we have suppressed the symbol $\displaystyle \lim_{s \to 0}$.
With the help of \eqref{eq:freehk} and \eqref{eq:beta}, 
equation \eqref{eq:diffexpand} becomes
\begin{align}
   \eqref{eq:diffexpand} 
    = \frac{i}{2} \frac{1}{(4\pi)^n} \int_0^1 dg \int dx' 
      \sum_{k,l,m} &  \frac{1}{m!} 
                      \frac{(k + m)!(l + m)!}
                      {(k + l + 2m +1)!}
                      s^{1 + k + l + m - n} 
      \notag \\
                 &\times \text{tr}\gamma_5 \gamma^\mu T^a 
                    \ti{b}_k(x,x') \Slash{A}^{\prime} b_l(x',x)
                       ( - \square)^m \delta (x - x' ) . 
      \label{eq:diffbeta}
\end{align}
Note that the terms higher than $0$-th order in $s$ vanish in the limit $s \to 0$, 
we find that the 
indices $k$, $l$, and $m$ of  the surviving terms on the right-hand side 
satisfy the condition
\begin{equation}
             1 + k + l + m - n \le 0 .        \label{eq:conditon_comon}
\end{equation}

Below, we work in two and four dimensions.

In two dimensions ($n = 1$), the condition \eqref{eq:conditon_comon} becomes  
\begin{equation}
                   k + l + m \le 0 ,
\end{equation}
which means that $k = l = m = 0$. Thus, equation \eqref{eq:diffbeta} reads
\begin{align}
   \langle J^{\mu a}(x) \rangle_{\textrm{cons}} 
      - \langle J^{\mu a}(x) \rangle_{\textrm{cov}} 
      &= 
      \frac{i}{8\pi} \int_0^1 dg\int dx' \text{tr} \gamma_5 \gamma^\mu T^a 
                  \ti{b}_0(x,x')\Slash{A}^\prime b_0(x',x) \delta (x - x')  
      \notag 
      \\
      &= 
      \frac{1}{4\pi} \vep^{\mu \nu} \text{tr} T^a A_\nu(x) ,
\label{eq:2dimdiff}
\end{align}
where we have used the coincidence limits $b_0(x,x) = \ti{b}_0(x,x) = \bm{1}$ 
(\eqref{eq:boundary3} and \eqref{eq:tilde-dagger}). 
This agrees with the previous results 
\cite{Bardeen_Zumino_gauge_gravity,Banerjee_Banerjee_Mitra}. 

In four dimensions ($n = 2$), the condition \eqref{eq:conditon_comon} becomes 
\begin{equation}
                       k + l + m - 1 \le 0 .
\end{equation}
The solutions of this condition are 
$(k,l,m) = (0,0,0),(1,0,0),(0,1,0),(0,0,1)$. 
Calculating the four terms corresponding 
to these solutions, we obtain
\begin{align}
  & \langle J^{\mu a}(x) \rangle_{\textrm{cons}} 
                - \langle J^{\mu a}(x) \rangle_{\textrm{cov}} 
  \notag 
  \\
  &= \frac{i}{2} \frac{1}{(4\pi)^2} \int_0^1 dg \int dx' \text{tr} \gamma_5\gamma^\mu T^a 
       \left( 
              \frac{1}{s} \ti{b}_0(x,x')\Slash{A}^\prime b_0(x',x) 
       \right.
  \notag 
  \\
  &\qquad 
       \left. 
             + \frac{1}{2} \ti{b}_1 (x,x') \Slash{A}^\prime b_0(x',x) 
             + \frac{1}{2} \ti{b}_0(x,x') \Slash{A}^\prime b_1 (x',x) 
             - \frac{1}{3!} \ti{b}_0(x,x' )\Slash{A}^{\prime}b_0(x' ,x) \square 
       \right) 
       \delta (x - x') 
  \notag 
  \\
  &= \frac{i}{2} \frac{1}{(4\pi)^2} \int_0^1 dg
       \left( \frac{1}{s} \text{tr} \gamma_5 \gamma^\mu T^a 
               [\ti{b}_0] \Slash{A}^\prime [b_0]
       \right.
  \notag 
  \\ 
  & \qquad 
       \left. 
            + \frac{1}{2} \text{tr} \gamma_5 \gamma^\mu T^a [\ti{b}_1] \Slash{A} [b_0] 
            + \frac{1}{2} \text{tr} \gamma_5 \gamma^\mu T^a [\ti{b}_0] \Slash{A}[b_1] 
            - \frac{1}{3!} \big[ \square^\prime \text{tr} \gamma_5 \gamma^\mu T^a 
                     \ti{b}_0 \Slash{A}^\prime b_0 \big]
        \right) , 
\label{eq:diffcoin}
\end{align}
where we have used Synge's symbol \cite{Synge} to denote  coincidence limits such as 
$[\partial_\alpha b_0] = \displaystyle \lim_{x' \to x} \partial_\alpha b_0(x,x')$.
Since $[b_0] = [\ti{b}_0] = \bm{1}$, 
the first term in the integrand of \eqref{eq:diffcoin} vanishes after taking 
the trace over the spinor indices.
With the help of 
coincidence limits \eqref{eq:boundary3}, \eqref{eq:siki3}, 
and \eqref{eq:tilde-dagger}, the second 
and third terms become 
\begin{align}
  &\frac{1}{2} \text{tr} \gamma_5 \gamma^\mu T^a 
           [\ti{b}_1] \Slash{A} [b_0] 
       = i(1+ g) \vep^{\mu \alpha \beta \gamma} 
             \text{tr} T^a A_\gamma \partial_\beta A_\alpha
         +(1 + g^2 ) \vep^{\mu \alpha \beta \gamma}
             \text{tr} T^a A_\alpha A_\beta A_\gamma , 
  \label{eq:third} 
  \\
  &\frac{1}{2} \text{tr} \gamma_5 \gamma^\mu T^a 
           [\ti{b}_0]\Slash{A}[b_1] 
       = i(1+g) \vep^{\mu \alpha \beta \gamma}
             \text{tr} A_\gamma T^a \partial_\beta  A_\alpha  
         +(1+g^2) \vep^{\mu \alpha \beta \gamma}
             \text{tr}T^a A_\alpha A_\beta  A_\gamma  .
  \label{eq:second}
\end{align}
The last term in \eqref{eq:diffcoin} can be calculated as follows. Note that
\begin{align}
  \square^\prime (\ti{b}_0 \Slash{A}^\prime b_0) 
     &= (\square^\prime \ti{b}_0) \Slash{A}^\prime b_0 
      + \ti{b}_0 \Slash{A}^\prime \square^\prime b_0 
      + \ti{b}_0 (\square^\prime \Slash{A}^\prime ) b_0 
  \notag 
  \\
     & \qquad 
       + 2(\partial_\alpha^\prime \ti{b}_0) 
          (\partial^{\prime \alpha} \Slash{A}^\prime)b_0 
       + 2 \ti{b}_0 (\partial_\alpha^\prime \Slash{A}^\prime) \partial^{\prime \alpha} b_0 
       + 2(\partial_\alpha^\prime \ti{b}_0) \Slash{A}^\prime \partial^{\prime \alpha } b_0 .
\label{eq:product}
\end{align}
The coincidence limit of \eqref{eq:product} can be evaluated by 
using \eqref{eq:boundary3}, \eqref{eq:siki1}, \eqref{eq:siki2},
and \eqref{eq:tilde-dagger}; thus, we obtain
\begin{equation}
  \big[ 
       \square' \text{tr} \gamma_5 \gamma^\mu T^a \ti{b}_0 \Slash{A}' b_0
  \big] 
  = 
  2i(1-g) \vep^{\mu \alpha \beta \gamma} \text{tr} \{T^a,A_\gamma \} 
                                                 \partial_\beta  A_\alpha  
  + 
  8(1-g) \vep^{\mu \alpha \beta \gamma} \text{tr} T^a A_\alpha A_\beta A_\gamma  .
\label{eq:first} 
\end{equation}
From these results, we finally obtain
\begin{align}
  &\langle J^{\mu a}(x) \rangle_{\textrm{cons}} 
    - \langle J^{\mu a}(x) \rangle_{\textrm{cov}} 
  \notag 
  \\
  &= 
    \frac{1}{24 \pi^2} \vep^{\mu \alpha \beta \gamma} \text{tr} \{T^a,A_\gamma \} 
                                \partial_\beta  A_\alpha  
  - \frac{i}{16\pi^2} \vep^{\mu \alpha \beta \gamma} \text{tr} 
                                T^a A_\alpha A_\beta  A_\gamma  , 
\label{eq:4dimdiff}
\end{align}
which agrees with the previous 
results \cite{Bardeen_Zumino_gauge_gravity,Banerjee_Banerjee_Mitra}.  

%
%
%
%
%
%
%
%
\section{Relationship between the covariant and consistent energy--momentum tensors}
\label{5}
In this section, we apply the arguments of 
Banerjee et al. \cite{Banerjee_Banerjee_Mitra}, as explained in Sect. \ref{2},  
to  gravitational anomalies \cite{Alvarez-Gaume_Witten}.
The vacuum expectation value of the energy--momentum tensor density 
$\langle eT^a _{\ \mu}(x) \rangle$ is expressed by 
the effective action $W[e_{b}^{\ \nu}]$:
\begin{equation}
  \langle eT^a _{\ \mu}(x) \rangle 
    = \frac{\delta}{\delta e_{a}^{\ \mu}(x)} W[e_{b}^{\ \nu}] , 
\label{eq:emtensoreff} 
\end{equation}
where $e_{a}^{\ \mu}$ is the vielbein field, 
$e = \text{det}\,e^a _{\ \mu}$ and $e^a _{\ \mu}$ is the 
inverse matrix of $e_{a}^{\ \mu}$.
Gravitational anomalies appear as non-zero values 
of $D^{\mu}\langle T^a _{\ \mu}(x) \rangle$ (Einstein anomaly) 
and/or 
$
\langle T_{[a b]}(x)\rangle
=\frac{1}{2}\langle T_{a b}(x)-T_{ba}(x)\rangle
$ (Lorentz anomaly).

In equation \eqref{eq:emtensoreff}, 
$\langle eT^a _{\ \mu}(x) \rangle$ is ill-defined because $W[e_{b}^{\ \nu}]$ 
is a divergent quantity. 
To treat the energy--momentum tensor $\langle eT^a _{\ \mu}(x) \rangle$ meaningfully, 
we should adopt an 
appropriate regularization,  
either 
consistent  or  covariant. 
The consistently regularized energy--momentum tensor 
$\langle eT^a _{\ \mu}(x) \rangle_{\text{cons}}$ 
is defined by the regularized effective action 
$W[e_{b}^{\ \nu}]_{\text{reg}}$ as 
\begin{equation}
  \langle eT^a _{\ \mu}(x) \rangle_{\text{cons}} 
       = \frac{\delta}{\delta e_{a}^{\ \mu}(x)}W[e_{b}^{\ \nu}]_{\text{reg}} . 
\label{eq:consemtensor}
\end{equation}
The covariant energy--momentum tensor $\langle eT^a _{\ \mu}(x) \rangle_{\text{cov}}$ 
is the expectation 
value of the energy--momentum tensor regularized covariantly with respect to both 
the general coordinate and local Lorentz 
transformations. 
These expectation values are functionals of $e_{a}^{\ \mu}(x)$.
When we need to pay attention to the functional property,
we use a symbol such as $\langle eT^a _{\ \mu}(x) \rangle_{\text{cov}}[e_{b}^{\ \nu}]$.

Now, we introduce a vielbein field, $e_{a}^{\ \mu}(t) = e_{a}^{\ \mu}(x,t)$,  
with one parameter $t$, which connects 
the original vielbein $e_{a}^{\ \mu}(x)$ 
to the flat space--time vielbein $\delta_{a}^{\mu}$ 
such that
\begin{align}
         &e_{a}^{\ \mu}(x,0) = \delta_{a}^{\mu} , 
         \\
          &e_{a}^{\ \mu}(x,1) = e_{a}^{\ \mu}(x) .
\end{align}
For example, we may adopt 
$e_{a}^{\ \mu}(t) = \de_{a}^{\mu} + t(e_{a}^{\ \mu}(x) - \de_{a}^{\mu})$ 
or 
$e_{a}^{\ \mu}(t) = (e^{tH})_{a}^{\ \mu}$ 
with the matrix $H_{a}^{\ \mu} = (\ln e)_{a}^{\ \mu}$.
We define a $t$-parametrized effective action $W_t$ by
\begin{equation}
W_t = W[e_{a}^{\ \mu}(t)] ,
\end{equation}
which reduces to the original effective action $W[e_{a}^{\ \mu}]$ if $t=1$.
We can express the effective action $W[e_{a}^{\ \mu}]$ by using $W_t$ as 
\begin{equation}
  W[e_{a}^{\ \mu}] = \int_0^1 dt\frac{\partial W_t}{\partial t} + W_{t = 0} .
\end{equation}
Note that the $t$-dependence of $W_t$ arises only through $e_{a}^{\ \mu}(t)$, 
we obtain
\begin{equation}
 W[e_{a}^{\ \mu}] 
      = \int_0^1 dt \int dx \,
             \frac{\delta W_t}{\delta e_{a}^{\ \mu}(x,t)}
             \frac{\partial e_{a}^{\ \mu}(x,t)}{\partial t} , 
   \label{eq:eff0}
\end{equation}
where we have dropped the $W_{t = 0}$ term,
 since it is an $e_{a}^{\ \mu}$-independent constant.
From definition \eqref{eq:emtensoreff}, we rewrite this equation as  
\begin{equation}
 W[e_{a}^{\ \mu}] 
          = \int_0^1 dt \int dx \langle eT^a _{\ \mu} \rangle^t
            \frac{\partial e_a^{\ \mu}(t)}{\partial t} ,   
 \label{eq:eff}
\end{equation}
where we have used the notation
\begin{equation}
 \langle eT^a_{\ \mu} \rangle^t = \langle eT^a_{\ \mu} \rangle [e_b^{\ \nu}(t)] .
\end{equation}
To construct a regularized effective action, $W_{\text{reg}}[e_{a}^{\ \mu}]$, 
we substitute the covariant energy--momentum tensor 
$
\langle eT^a _{\ \mu} \rangle^t_{\text{cov}} 
  = \langle eT^a _{\ \mu} \rangle_{\text{cov}}[e_{b}^{\ \nu}(t)]
$
for 
 $\langle eT^a _{\ \mu} \rangle^t$ 
on the right-hand side of equation \eqref{eq:eff}:
\begin{equation}
 W_{\text{reg}}[e_{a}^{\ \mu}]
          = \int_0^1 dt \int dx \langle eT^a _{\ \mu} \rangle^t_{\text{cov}}
            \frac{\partial e_{a}^{\ \mu}(t)}{\partial t} . 
  \label{eq:regeff}
\end{equation}
We can obtain a consistent energy--momentum tensor from the regularized effective 
action \eqref{eq:regeff}. 
Taking the variation of \eqref{eq:regeff} with respect to $e_{a}^{\ \mu}$, 
we obtain the following relationship between 
the covariant and consistent energy--momentum tensors:
\begin{align}
    &\int dx\langle eT^a _{\ \mu} \rangle_{\text{cons}}\delta e_{a}^{\ \mu} \notag 
    \\
    &= 
      \int_0^1 dt \int dx \langle eT^a _{\ \mu} \rangle^t_{\text{cov}}
                             \frac{\partial}{\partial t}\delta e_{a}^{\ \mu}(t) 
     + \int_0^1 dt \int dx \int dx'  
                      \frac{ \delta \langle eT^a _{\ \mu}\rangle_{\text{cov}}^t }
                           { \delta e_{b^{\prime}}^{\ \nu^{\prime}}(t) }
                      \delta e_{b'}^{\ \nu'}(t) 
                           \frac{\partial e_{a}^{\ \mu}(t)} {\partial t} 
    \notag \\
    &= 
     \int dx \langle eT^a _{\ \mu} \rangle_{\text{cov}}\delta e_{a}^{\ \mu} 
    \notag \\
    &\qquad \qquad 
      + \int_0^1 dt\int dx\int dx' 
            \left\{ 
                 \frac{\delta \langle eT^a _{\ \mu}\rangle_{\text{cov}}^t }
                      {\delta e_{b'}^{\ \nu'}(t)  } 
              -  \frac{\delta \langle e'T^{b'}_{\ \nu'}\rangle_{\text{cov}}^t}
                      {  \delta e_{a}^{\ \mu}(t)                            } 
            \right\} 
              \delta e_{b'}^{\ \nu'}(t) \frac{\partial e_{a}^{\ \mu}(t)}
                                          {\partial t               } , 
\label{eq:regeffcovemtensorfc1}
\end{align}     
where we have applied integration by parts to the first term in the second line 
and used the fact that the 
$t$-dependence of $\langle eT^a _{\ \mu} \rangle^t_{\text{cov}}$ arises only 
through $e_{a}^{\ \mu}(t)$. 
In \eqref{eq:regeffcovemtensorfc1}, 
primed indices denote those attached at the point $x' $ such as
\begin{equation}
    \langle e^{\prime}T^{b^{\prime}}_{\ \n^{\prime}} \rangle 
      = \langle e(x' )T^b_{\ \nu}(x' ) \rangle , 
    \quad 
    e_{b'}^{\ \nu'}(t) = e_{b}^{\ \nu}(x',t) .
\end{equation}

We emphasize that the ``functional curl'' of the covariant energy--momentum 
tensor appears in \eqref{eq:regeffcovemtensorfc1}. 
This curl vanishes only when the theory is not anomalous. 
In fact,  if the theory is anomaly free, the  regularized effective action is 
invariant under the general coordinate and 
local Lorentz transformations. 
In this case, 
the consistent energy--momentum tensor becomes covariant, and thus  
the 
covariant 
energy--momentum tensor 
satisfies the integrability condition, i.e., the condition of vanishing functional curl.
Conversely, if the functional curl of the covariant energy--momentum tensor is zero,  
the consistent energy--momentum tensor coincides with the covariant one, 
as seen from \eqref{eq:regeffcovemtensorfc1}. 
In this case, the covariant and consistent gravitational anomalies coincide with each other. 
The diagrammatic approach to the anomaly, however,  tells us that 
the leading terms of these anomalies differ by the Bose-symmetrization factor 
$1/(n+1)$ in $2n$ dimensions.  
This is true only when both anomalies are zero. 
Thus, the vanishing functional curl indicates an  anomaly-free theory.

The relationships between the covariant and consistent gravitational anomalies 
are derived immediately from 
 \eqref{eq:regeffcovemtensorfc1}.
 For example, if we adopt the parametrization 
 $e_{a}^{\ \mu}(t) = \de_{a}^{\mu} + t(e_{a}^{\ \mu}(x) - \de_{a}^{\mu})$,
equation \eqref{eq:regeffcovemtensorfc1} becomes
\begin{equation}
  \langle eT^a _{\ \mu} \rangle_{\text{cons}} 
     = \langle eT^a _{\ \mu} \rangle_{\text{cov}} 
     + \int_0^1 dt \int dx' t ( e_{b'}^{\ \nu'} - \delta_{b'}^{\nu'} )
        \left\{ 
         \frac{\delta \langle e'T^{b'}_{\ \nu'}\rangle_{\text{cov}}^t }
              {\delta e_{a}^{\ \mu}(t)                                   } 
       - \frac{\delta \langle eT^a _{\ \mu}\rangle_{\text{cov}}^t }
              {   \delta e_{b'}^{\ \nu'}(t)                             }
        \right\} .
  \label{eq:Tcons=Tcov+fc_a}
\end{equation}
Taking the covariant divergence of both sides, we obtain  a relationship 
between the covariant and consistent Einstein anomalies
\begin{equation}
    D^{\mu} \langle eT^a _{\ \mu} \rangle_{\text{cons}} 
    =
        D^{\mu}\langle eT^a _{\ \mu} \rangle_{\text{cov}} 
      + D^{\mu} \int_0^1  dt \int dx' 
           t (e_{b'}^{\ \nu'} - \delta_{b'}^{\nu'})
          \left\{
              \frac{\delta \langle e'T^{b'}_{\ \nu'} \rangle_{\text{cov}}^t}
                   {  \delta e_{a}^{\ \mu}(t)  } 
            - \frac{\delta \langle eT^a _{\ \mu}\rangle_{\text{cov}}^t}
                   { \delta e_{b'}^{\ \nu'}(t)}
          \right\} .
\end{equation}
The relationship between the Lorentz anomalies can be similarly obtained. 

%
%
%
%
%
%
%
\section{Summary and discussion}
\label{6}
In Sect. \ref{3}, 
we  evaluated the functional curl of the covariant current explicitly 
using the heat kernel method in arbitrary even dimensions. 
The result gives a direct proof of the delta-function-type behavior 
of the functional curl. 
Our explicit form of this curl leads immediately to the relationship between 
the covariant and consistent currents presented by  Bardeen and 
Zumino \cite{Bardeen_Zumino_gauge_gravity, Banerjee_Banerjee_Mitra}. 
In Sect. \ref{4}, we applied the heat kernel method 
to evaluate Osabe and Suzuki's results of the difference between the covariant and 
consistent currents \cite{Osabe_Suzuki}  in two and four dimensions.
The results are the same as previously reported 
\cite{Bardeen_Zumino_gauge_gravity, Banerjee_Banerjee_Mitra}. 
In Sect. \ref{5}, applying the arguments of Banerjee et al. 
\cite{Banerjee_Banerjee_Mitra} to gravitational anomalies, 
we have investigated the relationship between the covariant and consistent 
energy--momentum tensors. 
The relation is found to be expressed by the functional curl of 
the covariant energy--momentum tensor. 
%

The energy--momentum tensors considered in  Sect. \ref{5} have 
both 
Einstein  and Lorentz anomalies in general. As shown in Ref. 
\cite{Bardeen_Zumino_gauge_gravity}, 
these anomalies are not independent of each other. 
Moreover, using the regularization ambiguity, 
we can always choose the energy--momentum tensor to have either a vanishing Lorentz 
anomaly or a vanishing Einstein anomaly. 
From  the covariant regularization viewpoint, this is explained below. 

Given a covariantly regularized energy--momentum tensor, 
$\langle T_{\mu\nu} \rangle_\text{cov}$, 
we have in general both the Einstein anomaly, 
$D^\mu \langle T_{\mu\nu} \rangle_\text{cov}$, 
and the Lorentz anomaly,
$\langle T_{[\mu\nu]} \rangle_\text{cov}$.
Note 
that these covariant gravitational anomalies are local polynomials 
of the Riemann curvature (and its derivative for the Einstein anomaly). 
Because of the regularization ambiguity, we can add a finite, local, and covariant 
counterterm to 
$\langle T_{\mu\nu} \rangle_\text{cov}$ 
to obtain another covariantly regularized energy--momentum tensor. 
Adopting the Lorentz anomaly as a  counterterm, 
we can obtain a Lorentz-anomaly-free energy--momentum tensor, 
\begin{align}
\langle T_{\mu\nu} \rangle_\text{cov}^\text{pE}
  =
    \langle T_{\mu\nu} \rangle_\text{cov} - \langle T_{[\mu\nu]} \rangle_\text{cov}, 
\end{align}
which gives the pure covariant Einstein anomaly 
$
D^\mu \langle T_{\mu\nu} \rangle_\text{cov}^\text{pE}.
$
Since 
the energy--momentum tensor, 
$\langle T_{\mu\nu} \rangle_\text{cov}^\text{pE}$, 
given above is nothing but the symmetric part of 
$\langle T_{\mu\nu} \rangle_\text{cov}$, 
we can say that the pure covariant Einstein anomaly is 
the covariant divergence of the symmetric part of the covariant
energy--momentum tensor \cite{Alvarez-Gaume_Witten, F-T-Y, Endo_Takao, Fujikawa_Suzuki}.

We can also define a covariant energy--momentum tensor wiht a vanishing 
Einstein anomaly. It is known from 
\cite{Alvarez-Gaume_Witten, F-T-Y, Matsuki, Endo_Takao, Fujikawa_Suzuki, Peter}
that the pure  covariant Einstein anomaly has the form
\begin{align}
  D^\mu \langle T_{\mu\nu} \rangle_\text{cov}^\text{pE}
  = - D^\mu L_{\mu\nu},
\end{align}
where $L_{\mu\nu}$ is a local polynomial of Riemann curvature\footnote{%
   The quantity $L_{\mu\nu}$ is  related to axial $U(1)$ anomalies in $d+2$ dimensions 
  \cite{Alvarez-Gaume_Witten, F-T-Y, Matsuki, Endo_Takao, Fujikawa_Suzuki, Peter}. 
  }
and is anti-symmetric with respect to the indices $\mu$ and $\nu$. 
To obtain an Einstein-anomaly-free energy--momentum tensor, 
$\langle T_{\mu\nu} \rangle_\text{cov}^\text{pL}$, 
we may adopt  
$L_{\mu\nu}$ as a local counterterm to 
$\langle T_{\mu\nu} \rangle_\text{cov}^\text{pE}$: 
\begin{align}
    \langle T_{\mu\nu} \rangle_\text{cov}^\text{pL}
    =
    \langle T_{\mu\nu} \rangle_\text{cov}^\text{pE} + L_{\mu \nu}, 
\end{align}
which has vanishing Einstein anomaly:  
$D^\mu \langle T_{\mu\nu} \rangle_\text{cov}^\text{pL}=0$. 
Thus, the pure covariant Lorentz anomaly is given by
\begin{align}
    \langle T_{[\mu\nu]} \rangle_\text{cov}^\text{pL}
    =L_{\mu\nu}.
\end{align}

In Sect. \ref{5}, we have defined a regularized effective action 
 using the covariant energy--momentum tensor (equation \eqref{eq:regeff}). 
Since the covariant energy--momentum tensor retains the ambiguity of adding 
covariant local curvature and vielbein polynomials, corresponding 
ambiguity arises in the effective action \eqref{eq:regeff}. 
Then,  
one might wonder what kind of covariant energy--momentum tensor 
leads to the Lorentz-anomaly-free effective action, 
which is local Lorentz invariant but which does not have 
general coordinate invariance. 
It can be 
seen that the Lorentz-anomaly-free covariant energy--momentum tensor 
does not necessarily lead to a Lorentz-anomaly-free effective action. 
In fact, 
 for   spin-$1/2$ chiral fermions in two-dimensional space--time, 
an explicit calculation 
with the use of a symmetric covariant energy--momentum tensor 
$\langle eT_{\mu\nu} \rangle_\text{cov}^\text{pE}[e_b^{\ \lambda}(t)]$ 
to define the effective action \eqref{eq:regeff}
shows that 
the second term on the right-hand side of equation \eqref{eq:Tcons=Tcov+fc_a}
contributes to the consistent Lorentz anomaly. 
Thus, obtaining a Lorentz-anomaly-free (or Einstein-anomaly-free) 
consistent energy--momentum tensor 
is not yet straightforward in the context of \eqref{eq:regeffcovemtensorfc1}. 
Future work will aim to  clarify these points. 

%
%
%
%
%
%
%
%
\appendix
\section{Proof of \eqref{eq:freehk} using test functions}
\label{C}
In this appendix, we prove \eqref{eq:freehk}  using test functions. Namely, 
we give a proof of  equality
\begin{equation}
      \int d^{2n}xe^{x^2 /4s}f(x) 
                 = (4\pi s)^n \sum_{k = 0}^{\infty}
                    \int d^{2n}x\, f(x) \frac{( - s\square )^{k}}{k!}\delta (x) , 
\label{eq:prove}
\end{equation}
where $f(x)$ is an arbitrary test function. 
Changing the integration variables from $x^{\mu}$ to $\xi^{\mu}$
\begin{equation}
           \xi^{\mu} = \frac{x^{\mu}}{\sqrt{4s}} ,
\end{equation}
we express the left-hand side of \eqref{eq:prove} as
\begin{align}
   \text{LHS of } \eqref{eq:prove} 
      &= (4s)^n \int d^{2n}\xi \,e^{\xi^2}f(\sqrt{4s}\xi) 
      \notag 
      \\
      &= (4s)^n \sum_{l = 0}^{\infty}
               \frac{(4s)^{l/2}}{l!}
               \int d^{2n}\xi\, \xi^{\mu_1} \xi^{\mu_2} \cdots \xi^{\mu_l}
               \, e^{\xi^2} f_{,\mu_1 \mu_2 \cdots \mu_l}(0) , 
\label{eq:sec}
\end{align}
where we have taken the Taylor expansion of $f(\sqrt{4s}\xi)$ with respect to $\xi$
and 
$f_{,\mu_1 \mu_2 \cdots \mu_l}
  ={\partial^l f}/{\partial x^{\mu_l} \cdots \partial x^{\mu_2}  \partial x^{\mu_1}}
$. 
Owing to  formula
\begin{align}
  \int & d^{2n}\xi \,  
         \xi^{\mu_1} \xi^{\mu_2} \cdots \xi^{\mu_l}\,e^{\x^2 } 
       \notag \\
      &=\begin{cases}
         \dfrac{\pi^n}{2^k}(\delta^{\mu_1 \mu_2}\delta^{\mu_3 \mu_4}
                            \cdots \de^{\mu_{2k - 1}\mu_{2k}} 
        + \text{permutations} 
                          )  & \quad (l=2k) \\
          0 & \quad (l=2k+1)
          \end{cases},
\end{align}
\eqref{eq:sec} becomes
\begin{align}
 \eqref{eq:sec} 
      &= (4s)^n  \sum_{k = 0}^{\infty} \frac{(4s)^k}{(2k)!} \,
                \frac{\pi^n }{2^k}\, (2k-1)!! 
                                       \left( 
                                                (-\square)^k f 
                                      \right) (0)
      \notag 
      \\
      &= (4\pi s)^n \sum_{k = 0}^\infty 
                   \frac{\left( (-s\square)^k f \right)(0) }
                        {k!}  ,
\end{align}
which is equal to the right-hand side of \eqref{eq:prove}.

%
%
%
\section{The largest number of  gamma matrices included in $a_k(x,x' )$} 
\label{A}
%
The heat kernel 
$K(x,x' ;s) 
  = e^{ - s \not D ^2 } \delta (x - x' )
$ 
satisfies the 
differential equation
\begin{equation}
       - \frac{\partial K(x,x' ;s)}{\partial s} 
                 = \Slash{D}^2 K(x,x' ;s)             \label{eq:heq}
\end{equation}
and the boundary condition
\begin{equation}
      K(x,x'; 0) = \delta (x - x' ) .            \label{eq:boundary}
\end{equation}
We assume the following expansion of $K(x,x' ;s)$, 
\begin{equation}
        K(x,x' ;s) 
        = \frac{1}{(4\pi s)^n } e^{(x - x')^2/4s}
                 \sum_{k = 0}^\infty a_k(x,x') \, s^k . 
\end{equation}
From the boundary condition \eqref{eq:boundary}, we have 
\begin{equation}
            a_0(x,x) = \bm{1} .                    \label{eq:a0xx}
\end{equation}
Equation \eqref{eq:heq} leads to the following recurrence formulas for $a_k$'s:
\begin{align}
    &(x - x')^{\mu}D_{\mu}a_0 = 0 ,         
    \label{eq:recurrence0}   
    \\
    &(k + 1)a_{k+1} + (x - x')^\mu D_\mu a_{k+1} + D_\mu D^\mu a_k 
       + \frac{i}{2} \gamma^\mu \gamma^\nu F_{\mu \nu} a_k = 0 . 
       \qquad (k \ge 0) 
    \label{eq:recurrence1}
\end{align}

From equations \eqref{eq:a0xx} and \eqref{eq:recurrence0}, 
we  confirm that $a_0(x,x' )$ is the parallel-displacement matrix 
of gauge group \cite{DeWitt_HK}. 
Then, it is obvious that $a_0(x,x' )$ does not contain 
any gamma matrices $\gamma^\mu $. 
Equation \eqref{eq:recurrence1} 
shows that $a_{k+1}(x,x')$ has two more gamma matrices than $a_k(x,x')$,  
since $D_\mu$ has none. 
From these observations, 
we find that the largest number of  gamma matrices included in $a_k(x,x' )$ 
is equal to $2k$. 

In the coincidence limit $x'\to x$,  $a_k(x,x')$ still has at most 
$2k$ gamma matrices. 
In fact,  equations \eqref{eq:a0xx}, \eqref{eq:recurrence0},  
and \eqref{eq:recurrence1} lead us to 
\begin{equation}
    a_k(x,x) 
       = \frac{(-1)^{k}}{k!}
               \left( 
                      \frac{i}{2} \gamma^\mu \gamma^\nu F_{\mu \nu} 
               \right)^k 
       + \cdots , 
\label{eq:gammamax1}
\end{equation}
where the dots on the right-hand side express terms 
with lower power of gamma matrices. 

%
%
%
\section{Heat kernel appearing in Osabe and Suzuki's currents}
\label{B}
%
%
The heat kernel 
$K_g(x,x';s) 
  = \langle x|e^{-s \not D \not D_g} |x' \rangle
$ 
satisfies the  differential equation
\begin{equation}
   - \frac{\partial K_g(x,x';s)}{\partial s} = \Slash{D} \Slash{D}_g K_g(x,x';s) 
\label{eq:heq2}
\end{equation}
and the boundary condition
\begin{equation}
       K_g(x,x';0) = \delta (x - x') .             \label{eq:boundary2}
\end{equation}
We assume the following expansion of $K_g(x,x' ;s)$
\begin{equation}
      K_g(x,x' ;s) 
           =  \frac{1}{(4\pi s)^n}  e^{(x - x')^2/4s}
                              \sum_{k = 0}^\infty b_k(x,x' )\,s^k . 
\label{eq:K-expansion}
\end{equation}
From the boundary condition \eqref{eq:boundary2}, we have
\begin{equation}
          [b_0] = \bm{1} ,         \label{eq:boundary3}
\end{equation}
where we have used Synge's symbol to denote the coincidence limit 
$[f] = f(x,x)$.
Equation \eqref{eq:heq2} leads to the following recurrence formulas for
the $b_k$'s:
\begin{align}
          &(x - x')^\mu \mathscr{D}_\mu b_0 = 0 , 
          \label{eq:zenka1} 
          \\
          & (k+1) b_{k+1} + (x-x')^\mu \mathscr{D}_\mu b_{k+1} 
                          + (\mathscr{D}_\mu \mathscr{D}^\mu  +  P) b_k = 0 , 
          \qquad (k \ge 0) 
          \label{eq:zenka2} 
\end{align}
where we have used the  equation
\begin{align}
       & \Slash{D} \Slash{D}_g = \mathscr{D}_\mu \mathscr{D}^\mu +  P 
\end{align}
with
\begin{align}
   & \mathscr{D}_\mu = \partial_\mu + \mathscr{A}_\mu ,
   \\
   & \mathscr{A}_\mu 
         = \frac{i}{2}(1+g) A_\mu 
         + \frac{i}{2}(1-g) \gamma_{\alpha \mu} A^\alpha , 
   \\
   & P = - \frac{i}{2}(1-g) \partial_\alpha A^\alpha 
         - \frac{1}{2}(n-1)(1-g)^2 A_\alpha A^\alpha 
   \notag 
   \\
           &\qquad \qquad 
           + \frac{i}{2} \gamma^{\alpha \beta}
                  \left\{ 
                       (1+g)\partial_\alpha A_\beta   
                          +i \left( 
                                   (n-1)g^2 -2(n-2)g+ n - 1 
                             \right)
                                    A_\alpha A_\beta  
                  \right\}
\end{align}
and
$\gamma^{\mu \nu} = \frac{1}{2}[\gamma^\mu, \gamma^\nu]$.
From recurrence formulas \eqref{eq:zenka1} and \eqref{eq:zenka2}, 
together with \eqref{eq:boundary3}, we obtain
the following coincidence limits \cite{DeWitt_HK}:
\begin{align}
     &[\mathscr{D}_\mu b_0] = 0 ,    \label{eq:limit1} 
     \\
     &[\mathscr{D}_\mu \mathscr{D}_\nu b_0] 
            = \frac{1}{2} \mathscr{F}_{\mu \nu} ,      \label{eq:limit2} 
     \\
     &[b_1 ] = -  P ,                     \label{eq:limit3}    
\end{align}
where 
$
    \mathscr{F}_{\mu \nu} 
         = \partial_\mu \mathscr{A}_\nu 
         - \partial_\nu \mathscr{A}_\mu 
         + [\mathscr{A}_\mu , \mathscr{A}_\nu]
$.
From these equations, we obtain
\begin{align}
    &[\partial_\alpha b_0] 
      = - \frac{i}{2}(1+g) A_\alpha 
        - \frac{i}{2}(1-g) \gamma_{\beta \alpha}A^{\beta} , 
    \label{eq:siki1} 
    \\
    &[\square b_0] 
       = - \frac{i}{2}(1+g) \partial_\alpha A^\alpha 
         + \frac{1}{2} \left( (n-1)g^2  -2ng + n-1 \right) A_\alpha A^\alpha
   \notag 
   \\
   & \qquad \qquad \qquad 
         + \frac{i}{2} \gamma^{\alpha \beta}
                \left( (1-g) \partial_\alpha A_\beta 
                      - i(n-1)(1-g)^2 A_\alpha A_\beta  
                \right) ,
   \label{eq:siki2} 
   \\
   &[b_1 ] = \frac{i}{2}(1-g) \partial_\alpha A^\alpha 
           + \frac{1}{2}(n-1)(1-g)^2 A_\alpha A^\alpha 
   \notag 
   \\
   &\qquad \qquad 
           - \frac{i}{2} \gamma^{\alpha \beta}
                 \left\{ (1+g) \partial_\alpha A_\beta   
                        +i \left( 
                                  (n-1)g^2  - 2(n-2)g + n-1 
                           \right) 
                                  A_\alpha A_\beta  
                 \right\}  .
   \label{eq:siki3}
\end{align}


\begin{thebibliography}{99}
\bibitem{Wess_Zumino_cons}
  J.~Wess and B.~Zumino, Phys. Lett. B  {{\bf 37}, 95 (1971)}.  
%
\bibitem{Bardeen_Zumino_gauge_gravity}
  W.~A.~Bardeen and  B.~Zumino, Nucl. Phys. B
 {{\bf 244}, 421 (1984)}. 
%
\bibitem{Banerjee_Banerjee_Mitra}
  H. Banerjee,  R. Banerjee  and  P. Mitra,  Z. Phys. C
{{\bf 32}, 445 (1986)}. 
%
\bibitem{Banerjee_Banerjee}
  H. Banerjee and R. Banerjee,  Phys. Lett. B
{{\bf 174}, 313 (1986)}.
%
\bibitem{Banerjee_Banerjee_point_splitting}
  R. Banerjee and H. Banerjee,  Z. Phys. C
 {{\bf 39}, 89 (1988)}.
%
\bibitem{Fujikawa_Suzuki}
  K. Fujikawa and H. Suzuki,
  \textit{Path Integrals and Quantum Anomalies} (Oxford Univ. Press, 2004).  
%
\bibitem{Suzuki_SUSY}
  Y. Ohshima, K. Okuyama, H. Suzuki and H. Yasuta,
  Phys. Lett. B {{\bf 457}, 291 (1999)}.
%
\bibitem{Osabe_Suzuki}
  S. Osabe and H. Suzuki,  Int. J. Mod. Phys. A 
{{\bf 21}{\bf 9}, 3377 (1994)}.
%
\bibitem{Grisaru_Penati}
  Marcus T. Grisaru and S. Penati,  Phys. Lett. B 
{{\bf 504}, 89 (2001)}.
%
\bibitem{Dunne_Trugenberger}
  Gerald V. Dunne and Carlo A. Trugenberger,  Ann. Phys.
 {{\bf 204}, 281 (1990)}.
%
\bibitem{Kelnhofer}
  G. Kelnhofer,  Z. Phys. C
 {{\bf 52}, 89 (1991)}.
%
\bibitem{Z_Qiu_H_Ren}
  Z. Qiu and H. Ren  Phys. Rev. D
 {{\bf 38}, 2530 (1988)}.
%
\bibitem{DeWitt_HK}
  B. S. DeWitt,
  \textit{Dynamical Theory of Groups and Fields} (Gordon and Breach, 1965). 
%
\bibitem{Alvarez-Gaume_Witten}
  L. Alvarez-Gaum\'e and E. Witten, Nucl. Phys. B 
 {{\bf 239}, 269 (1984)}. 
\bibitem{Zumino_Y_S}
  B. Zumino, Y. -S. Wu, and A. Zee,  Nucl. Phys. B
 {{\bf 239}, 477 (1984)}. 
%
%
\bibitem{Synge}
  J. L. Synge,
  \textit{Relativity: The General Theory} (North-Holland Publishing, 1960).
%
\bibitem{F-T-Y}
  K. Fujikawa, M. Tomiya and O. Yasuda, Z. Phys. C {\bf 28}, 289 (1985).
%
\bibitem{Endo_Takao}
  R. Endo and M. Takao, Prog. Theor. Phys. {\bf 78}, 440 (1987).
%
\bibitem{Matsuki}
  T. Matsuki, Prog. Theor. Phys. {\bf 75}, 461 (1986).
%
\bibitem{Peter}
  F. Bastianelli and P. van Nieuwenhuizen, 
  \textit{Path Integrals and Anomalies in Curved Space} 
  (Cambridge Univ. Press, 2006).%
%
\end{thebibliography}
\end{document}